\documentclass[pre,amsmath,amssymb,nofootinbib,showpacs,onecolumn,notitlepage]{revtex4-2}
\usepackage{graphicx}

\newcommand{\dd}{\partial}
\newcommand{\rd}{\mathrm{d}}

\newcommand{\eps}{\epsilon}

\newcommand{\beq}{\begin{equation}}
\newcommand{\eeq}{\end{equation}}

\DeclareMathOperator{\arctanh}{arctanh}
\DeclareMathOperator{\arcsinh}{arcsinh}

\newcommand{\n}{n}
\newcommand{\npa}{t}
\newcommand{\g}{\mathcal{F}}

\begin{document}

\title{The $\theta$-formulation of the 2D elastica -- Buckling and boundary layer theory.}

\author{
Gregory Kozyreff$^{1}$, Emmanuel Si\'efert$^{2}$, Basile Radisson$^{2}$ and Fabian Brau$^{2}$}

\address{$^{1}$Optique Nonlin\'eaire Th\'eorique, Universit\'e libre de Bruxelles (U.L.B.), CP 231, Campus de la Plaine, 1050 Bruxelles, Belgium.\\
$^{2}$Nonlinear Physical Chemistry Unit, Universit\'e libre de Bruxelles (U.L.B.), CP 231, Campus de la Plaine, 1050 Bruxelles, Belgium.}




\begin{abstract}
The equations of a planar elastica under pressure can be rewritten in a useful form by parametrising the variables in terms of the local orientation angle, $\theta$, instead of the arc length. This ``$\theta$-formulation'' lends itself to a particularly easy boundary layer analysis in the limit of weak bending stiffness. Within this parameterization, boundary layers  are  located at inflexion points, where $\theta$ is extremum, and they connect regions of low and large curvature. A simple composite solution is derived without resorting to elliptic functions and integrals. This approximation can be used as an elementary building block to describe complex shapes. Applying this theory to the study of an elastic ring under uniform pressure and subject to a set of point forces, we discover a snapping instability. This instability is confirmed by numerical simulations. Finally, we carry out experiments and find good agreement of the theory with the experimental shape of the deformed elastica. 
\end{abstract}
\maketitle

\section{Introduction}
The problem of finding the shape of an elastica under various loads and constraints appears in many contexts of the applied sciences and at a wide variety of scales, from biological cells to suspended bridges. After centuries of analysis~\cite{Love1927,Levien2008}, one may consider it to be  solved, at least as far as two-dimensional deformations is concerned. 
A general solution of the associated differential equations is quoted in Love's treatise~\cite{Love1927} and 
%
further developments abound in the literature on its basis~\cite{Linner1998,Pozrikidis2002,Vassilev2008,Djondjorov2011,Mora2012,Batista2014,Marple2015,Giomi2019,Barbieri2020}. It is thus often possible to explicitly write the coordinate of any point of the elastic curve as a function of arc length in a wide range of situations. A weakness of the known solutions, however, is that they involve several elliptic functions, which can themselves appear as arguments of inverse trigonometric functions. Furthermore, the parameters of these elliptic functions are themselves functions of the roots of a fourth order polynomial in which the mechanical parameters of the system appear. Hence, despite being undeniably useful, the exact solution can be quite hard to analyse by inspection. Neither does it simplify in the limit of a small bending stiffness or  long elastica. In response to this issue, the purpose of this paper is to put forward another way of parameterising the elastic curve that may lead to great simplification. Very simply, we propose to express the curvature and the internal forces not as a function of arc length but as  a function of the local angle, $\theta$, made by the tangent of the elastic curve with a reference direction. We thus call it the $\theta$-formulation of the elastica. It is no more than the application of an old trick to reduce the order of a differential system~\cite{BObook}, which has been used before~\cite{Shao-guang1996,Shao-guang1997}. What hasn't been appreciated so far, however, is its great advantage for asymptotic analysis, \textit{i.e.}, in the treatment of boundary layers.

After deriving this $\theta$-formulation, we  first show how it gives way to an economical resolution of some classical problems. Next, we exploit the limit of a long elastica or, equivalently, of weak bending stiffness. We assume that a net pressure acts on one side of the elastica and allow that a set of point forces be applied. In particular, we  devise a general framework to treat closed rings that are either compressed or inflated while being  either locally stretched or pinched. We also carry out experiments of such elastic rings and find that the asymptotic approximation faithfully reproduce observations. In the case of pinching with prescribed forces, we identify a snapping instability, past which the ring collapses to self-contact. Finally, we indicate how the same technique could be applied to more general problems.

\section{Derivation of the $\theta$-formulation}

The differential equations governing the shape of an elastica in a plane can be written as
\begin{align}
\dd_s\n&=p-\kappa\npa, \label{eq:nperp1}\\
\dd_s\npa&=\kappa\n, \label{eq:npara1}\\
B\dd_s\kappa&=-\n,  \label{eq:kappa1}
\end{align}
where $\kappa$, $\npa$, and $\n$ are the curvature, tension and shear force, respectively, $s$ is the arc length, and $B$ is the bending stiffness. Given the 1D nature of the elastica, the pressure $p$ appearing in Eq.~(\ref{eq:nperp1}) has dimensions of Nm$^{-1}$. Knowing $\kappa$, the local angle with respect to the horizontal direction is deduced from
\beq
\dd_s\theta=\kappa   \label{eq:theta1}
\eeq
and the cartesian coordinates along the elastic curve are solution of $\dd_sx=\cos \theta$,
$\dd_sy=\sin\theta$.
Multiplying  Eq.~(\ref{eq:kappa1}) by $\kappa$ and summing with (\ref{eq:npara1}), one  easily finds that
\beq
B\frac{\kappa^2}2+\npa=H,
\label{eq:defH}
\eeq
where $H$ is a constant (it is equivalent to the constant $\mu$ appearing in various papers~\cite{Arreaga2002,Vassilev2008,Djondjorov2011} and $C$ in~\cite{Marple2015}). This allows one to eliminate the variable $\npa$ and reduce the order of the differential system~(\ref{eq:nperp1})-(\ref{eq:theta1}) by one. Note that if we write $H$ as $H=B\kappa_0^2/2$, then a state with $\kappa=\kappa_0$ for all $s$ corresponds to $\npa=\n=p=0$. In that case, $H$ measures the curvature of a stress-free elastic ring. On the other hand, $\kappa=0$ everywhere corresponds to the state of  a straight elastica with  $\n=p=0$ and a uniform tension $\npa=H$. Let us also note that, if $\npa>0$  anywhere in the elastica, then $H>0$. Stated differently, the elastica must be everywhere under axial compression for $H$ to be negative.

Using (\ref{eq:defH}) to eliminate $\npa$, Eq.~(\ref{eq:nperp1}) yields
\beq
\dd_s\n=p-\kappa\left(H-B\kappa^2/2\right),
\label{eq:nperp2}
\eeq
which, together with Eqs.~(\ref{eq:kappa1}) and (\ref{eq:theta1}), yields a third-order differential system. A further reduction of order is achieved by expressing $\n$ and $\kappa$ as functions of $\theta$. Firstly, writing $\n=\n\left(\theta(s)\right)$, we have $\dd_s\n=\kappa\dd_\theta\n$. Hence, Eq.~(\ref{eq:nperp2}) becomes
\beq
\dd_\theta\n=\frac{p}{\kappa}- H+B\frac{\kappa^2}{2}.
\label{eq:nperp3}
\eeq
Secondly, if we write
\beq
\frac{\kappa^2}2=q\left(\theta(s)\right),
\eeq
we obtain, by differentiating each side with respect to $s$, 
\beq
\dd_s\kappa=\dd_\theta q.
\eeq
Hence,  Eq.~(\ref{eq:kappa1}) yields
\beq
\n=-B\dd_\theta q.
\label{eq:nperp2bis}
\eeq
Eventually, Eq.~(\ref{eq:nperp3}) becomes
\beq
\text{
\framebox{
$B\left(\dd^2_{\theta}q+q\right)+\frac{p}{\kappa}= H.$
}
}
\label{eq:q1}
\eeq
This equation, which we presently call the $\theta$-formulation, is not particularly useful from a numerical point of view but can lead to significant analytical progress, particularly in the limit of small and large curvature. Here, ``small'' and ``large'' curvature respectively refer to situations where  $p/\kappa$ or $B (\dd^2_{\theta}q+q )$ dominate the left hand side, respectively. Evidently, both limits considerably simplify the resolution of Eq.~(\ref{eq:q1}). To make this statement more systematic, let us rescale $\kappa$ (and therefore $q$) as
\begin{align}
\kappa&=\frac{p}{H}\,K, &q&=\left(\frac{p}{H}\right)^2Q,
\label{rescale}
\end{align}
provided that $p$ is constant and not zero. With this scaling, we still have $K^2/2=Q$ and Eq.~(\ref{eq:q1}) becomes
\beq
\eps^2\left(\dd^2_{\theta}Q+Q\right)+\frac{1}{K}= 1,
\label{eq:q2}
\eeq
where
\beq
\eps^2=\frac{Bp^2}{H^3}.
\label{def:eps}
\eeq

Viewing  $x$ and $y$ as functions of $\theta$, we have $\dd_s (x,y)=\kappa \dd_\theta (x,y)$. Hence, given $\kappa(\theta)$, the shape of the elastica is obtained by integrating
\begin{align}
\dd_\theta x&=\frac{\cos\theta}{\kappa(\theta)},
&\dd_\theta y&=\frac{\sin\theta}{\kappa(\theta)},
\label{elastica:xy}
\end{align}
or, in a more compact way,
\begin{align}
\dd_\theta z&=\frac{e^{i\theta}}{\kappa(\theta)},
&z&=x +i y.
\label{elastica:z}
\end{align}
The advantage of a complex representation is that translation by a vector $(x_0,y_0)$ amount to adding of $x_0+iy_0$ to $z$, while rotation by an angle $\alpha$ is achieved by a multiplication by $e^{i\alpha}$. This is useful for constructing elastic curves as pieces of solutions of the above differential equation joined together.
Furthermore, from Eqs.~(\ref{eq:defH})  and (\ref{eq:nperp2bis}), we have 
\begin{align}
\npa&=H\left(1-\eps^2Q\right), &\n&=-\eps^2 H \,\dd_\theta Q.
\label{H_is_npa}
\end{align}

So far, our analysis has been exact. In Secs.~\ref{sec:Euler} and \ref{sec:buckling}, we show how the above formulation can be useful to treat  some classical problems. Next, in Sec.~\ref{sec:hook}, we  derive asymptotic approximations of $K$ (hence, of $Q$) in the small-$\eps$ limit using the method of matched asymptotic expansions~\cite{BObook}. More specifically, we  derive simple formulas describing how the elastica switches between a small positive curvature to a large negative one. As a function of $\theta$, $K$ displays boundary layers near extremal value of $\theta$, which are inflexion points. By convention here, the positivity of the curvature is defined relative to the sign of the pressure: a positive curvature is understood here as being of the same sign as $p/H$. With this convention in mind, we may treat in one go elastic rings that are subjected to an inflating pressure or to an external compression.

\section{Application 0: Euler's elastica revisited}\label{sec:Euler}
Before investigating boundary layers, we wish to highlight the usefulness of writing Eq.~(\ref{eq:q1}) for the elastica problem posed and solved by the Bernoullis and Euler~\cite{Levien2008}. In that case, $p=0$, so the rescaling (\ref{rescale}) does not apply. However,  Eq.~(\ref{eq:q1}) for the curvature is now turned into a linear equation of the simplest kind. Its solution is simply and exactly
\begin{align}
q(\theta)&=\frac HB\left(1+c\cos\theta\right), 
&\kappa(\theta)&=\pm \sqrt{\frac {H}B}\sqrt{2\left(1+c\cos\theta\right)}.
\label{euler:q}
\end{align}
where $c$ is an arbitrary constant and $\theta=0$ defines a direction of reference. Specifying this direction of reference is equivalent to imposing the second constant of integration of the equation for $q$. With $x$ and $y$ respectively defined as the cartesian coordinates along and perpendicular to the reference direction, they satisfy Eqs.~(\ref{elastica:xy}). 
They yield 
\begin{align}
x(\theta)&=\sqrt{\frac{2B}{H}} \Bigg\{x_0 
\pm\frac{\sqrt{1+c}}{c}\left[E\left(\frac\theta2\Big|\frac{2c}{1+c}\right)-\frac1{1+c} F\left(\frac\theta2\Big|\frac{2c}{1+c}\right)\right]\Bigg\}
\label{euler:x},\\
y(\theta)&=\sqrt{\frac{2B}{H}} \left[y_0\mp\frac1c\left(\sqrt{1+c\cos\theta}-1\right)\right].
\label{euler:y}
\end{align}
Above, $x_0$ and $y_0$ are arbitrary constants, while the function $F(\phi|m)$ and $E(\phi|m)$ are the incomplete elliptic integrals of the first and second kinds, respectively~\cite{Abramowitz}. Note that, up to a constant vertical shift, the above expression makes it directly apparent that $y\propto\kappa$. 

\begin{figure}
\centering
\includegraphics[width=.8\textwidth]{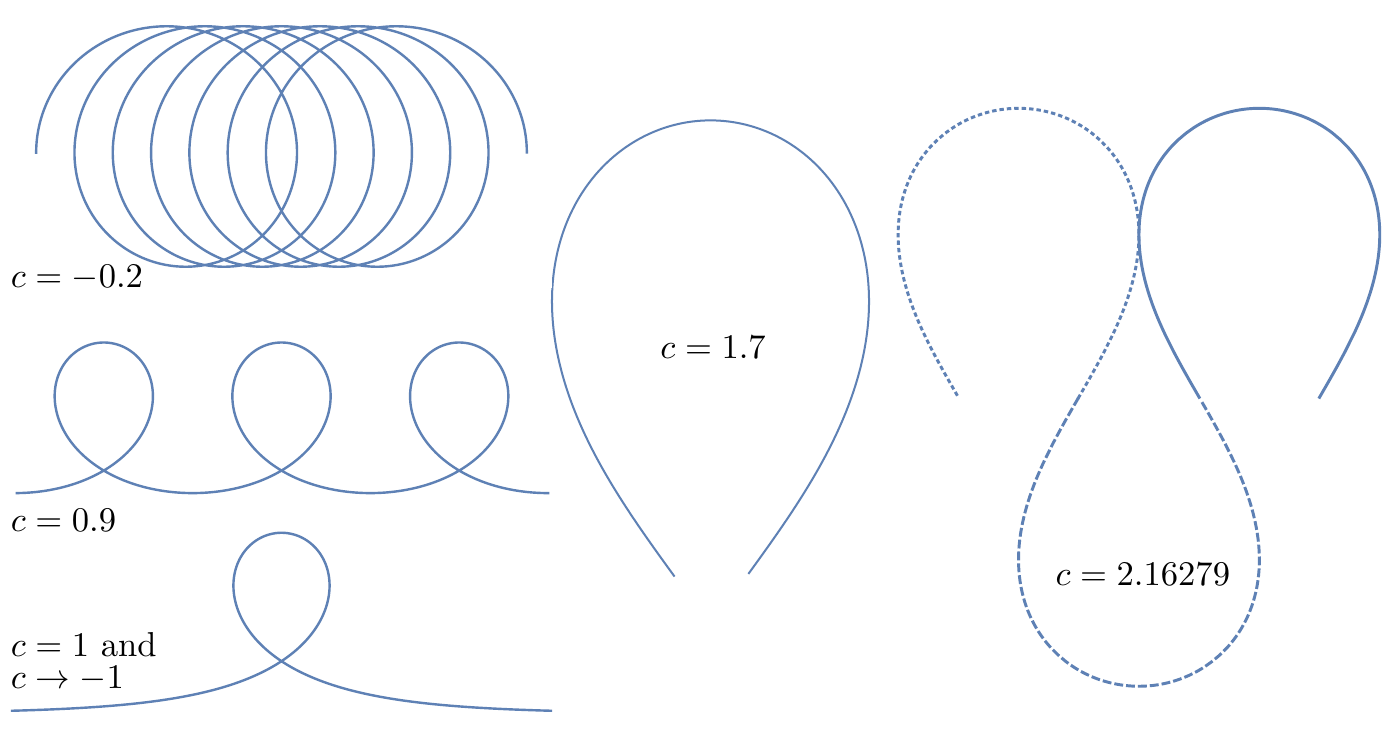}
\caption{Some examples of classic elastic curves ($p=0$) obtained for different values of $c$ in Eqs.~(\ref{euler:y}) and (\ref{euler:x}). With $c>1$, the curve contains inflexion points, at which copies of  Eqs.~(\ref{euler:y}) and (\ref{euler:x}) with different integration constants $c_{x,y}$ can be smoothly joined. An example is given with $c=2.16279$, value above which such composite curves without self-crossing can be constructed. }\label{fig:Euler}
\end{figure}

Eqs.~(\ref{euler:x}) and (\ref{euler:y}) are an alternative parametric expression of the classical elastica curves to what is found, \textit{e.g.} in Refs.~\cite{Love1927,Vassilev2008}. It appears that the $\theta$-formulation affords a somewhat more economical and direct route to the solution, as well as a minimal use of elliptic functions. The shape of the elastica is presently determined by the parameter $c$, while $H$ simply contributes to a scale factor, see Fig.~\ref{fig:Euler}. It is required that $-1<c<\infty$, with $c=0$ obviously corresponding to a perfect circle ($\kappa$ is constant in that case). The value $c=1$ corresponds to a single loop and so does the limiting case $c\to-1$. For $c>1$, $q(\theta)$ vanishes for some values of $\theta$, indicating inflexion points. At these, different instances of Eqs.~(\ref{euler:x}) and (\ref{euler:y}) can be used with distinct values of $c_x$ and $c_y$ and possibly a  change of the sign of $\kappa$ to construct a smooth elastica curve in a piecewise manner. As an illustration of this last point, we draw in Fig.~\ref{fig:Euler} the elastic curve for $c=2.16279$. This is the limiting value for the existence of self-crossing.

It will be useful in what follows to detail further the one-loop solution with $c=-1$, also known as syntractix curve~\cite[p. 277]{Salmon1873}. Then $q=2(H/B)\sin^2\theta/2$, $\kappa=\pm2\sqrt{H/B}\sin\theta/2$ and the equation for $x$ and $y$ can be integrated as
\begin{align}
x&=x_0\pm\sqrt{\frac BH}\left(2\cos\frac\theta2+\ln\left|4\tan\frac\theta4\right|\right),
&y&=y_0\pm2\sqrt{\frac BH} \sin\frac\theta2 .
\end{align}
In complex notation ($z=x+iy$), this is
\beq
z=z_0\pm \sqrt{\frac BH} \left(2 e^{i\theta/2}+\ln\left|4\tan\frac\theta4\right|\right).
\label{syntractrix}
\eeq
(The factor $4$ inside the logarithm above is introduced for convenience, in view of studying this expression in the limit where the argument $\theta\to0$.)

\section{Application 1: buckling instability of a compressed elastic ring}\label{sec:buckling}
To further demonstrate the interest of Eq.~(\ref{eq:q1}) or, equivalently, Eq.~(\ref{eq:q2}), let us consider a circular ring subject to uniform inward pressure and compute the buckling threshold at which it ceases to be circular. We thus look for a solution of the form
\begin{align}
K&\sim K_0+\delta \cos(n\theta)+\dotsb, 
&Q&\sim K_0^2/2+\delta K_0\cos\left(n\theta\right)+\dotsb,
\end{align}
where $\delta\ll1$ and $n$ is a positive integer. Above, $K_0$ is the rescaled curvature prior to buckling. Substituting this ansatz in  Eq.~(\ref{eq:q2}), we get ($\eps^2$ is not necessarily small, here, and not necessarily positive either)
\beq
\eps^2 \frac{K_0^2}{2}+\frac1{K_0}-1+\delta\left[\eps^2K_0\left(-n^2+1\right)-\frac{1}{K_0^2}\right]+O\left(\delta^2\right)=0.
\eeq
This yields two equations for the two unknowns $K_0$ and $\eps^2$:
\begin{align}
\eps^2 \frac{K_0^2}{2}+\frac1{K_0}&=1, & 
\eps^2K_0\left(-n^2+1\right)-\frac{1}{K_0^2}&=0.\label{buckling2}
\end{align}
Given that $n$ was defined as a positive integer, it directly follows from the second equation above that $n\neq1$ and, hence, $n\geq2$. Given the ring radius $R=1/\kappa$ prior to buckling, we have from (\ref{rescale}) that $K_0=H/(pR)$. Substituting this expression in the second of Eqs.~(\ref{buckling2}) and using  Eq.~(\ref{def:eps}) to eliminate $\eps^2$, one directly obtains  the classical buckling pressure derived by Halphen~\cite{Halphen1888} (see also~\cite{Arreaga2002}):
\beq
-p=\frac{B}{R^3} \left(n^2-1\right).
\eeq
Meanwhile, from the first of Eqs.~(\ref{buckling2}), we obtain
\beq
H=\frac{B }{ 2 R^2}  +pR = \frac{B }{R^2}\left(\frac32-n^2\right),
\eeq
which completes the description. Note that $H<0$, and hence $\eps^2<0$ in the present situation. We thus see that the $\theta$-formulation provides a particularly short route to the answer of this classical problem. Incidentally, by Eq.~(\ref{eq:defH}), the  compression along the elastica at buckling threshold is $-\npa=(n^2-1)B/R^2$. In terms of  its  half-perimeter  $L$, this is $ \pi^2(n^2-1)B/L^2$, to be compared with the critical buckling load of compression  $\pi^2 n^2B/L^2$ for a hinged straight beam of the same length~\cite{Howell09}. Recently, the dynamical aspects of this buckling instability have been examined both theoretically and experimentally~\cite{Box2020,Kodio2020}. Interestingly, they used the analogy with a straight beam under effective compression as a toy model for their experiment and, in~\cite{Kodio2020}, started out their analysis by introducing the dimensionless parameter $P=pR^3/B$. We will return to this adimensional parameter in \S~\ref{sec:peanut}.

\section{Application 2: small-$\eps$ hook shape}\label{sec:hook}
From now on, we assume a small bending stiffness, in the sense that $0<\eps\ll1$, and exploit this limit using boundary layer analysis. Implicitly, we restrict our attention to $H>0$, which includes all situations where $t>0$ at least at one point in the elastica. We first consider an elementary curve made up of two segments, one  with small curvature, the other with a large negative curvature. The resulting hook-curve will be used in latter sections as a building block to construct more complex shapes, such as the experimental shape that will be produced in \S~\ref{sec:peanut}, see Figs~\ref{fig:peanuta} and \ref{fig:trefoil1}. 
\subsection{Region of small curvature}\label{sec:smallcurv}
In Eq.~(\ref{eq:q2}), let us assume that $K=O(1)$. Then, letting $\eps\to0$, we obtain
\beq
K\sim K_I=1+O(\eps^2).
\label{eq:K_I}
\eeq
In view of Eq.~(\ref{rescale}), the above expression means $p\sim\kappa H$. Next, $H\sim\npa$, see Eq.~(\ref{H_is_npa}). Therefore, $K\sim1$ simply expresses that
\beq
p\sim \kappa \npa,
\eeq
which is nothing but the classical force balance equation of a membrane. Furthermore, since the curvature is approximately constant, the elastica is locally circular and we may infer without further calculation that,
\begin{equation}
\begin{pmatrix}x \\y \end{pmatrix}
\sim \begin{pmatrix}x_I \\y_I \end{pmatrix}
\sim \begin{pmatrix}  x_0\\   y_0\end{pmatrix}+
\frac{  H}{p}
\begin{pmatrix}  \sin\theta\\   -\cos\theta
\end{pmatrix}+O(\eps^2),
\label{smallcurv:shape}
\end{equation}
where $x_0$ and $y_0$ are constants of integration. In complex notation, this is
\beq
z\sim z_I\sim z_0-\frac{iH}{p} e^{i\theta}.
\label{eq:z_I}
\eeq

\subsection{Region of large absolute curvature}\label{sec:largecurv}
Where the curvature is large, let us assume that
\begin{align}
K&=\eps^{-1}K_{II}, &Q&=\eps^{-2}Q_{II}, 
&\text{(recall that $Q=K^2/2$)}
\label{scaling:reg_II}
\end{align}
where $K_{II}, Q_{II}=O(1)$. This leads to a different balance between terms in
Eq.~(\ref{eq:q2}), which becomes
\beq
\dd^2_{\theta}Q_{II}+Q_{II} +\frac{\eps}{K_{II}}= 1,
\label{eq:q3}
\eeq
Neglecting  the $O(\eps)$ term above, we directly obtain
\begin{align}
Q_{II} &\sim Q_{II}^{(0)}=  1+c\cos\left(\theta-\alpha\right),
&\to K_{II}&\sim K_{II}^{(0)}= -\sqrt{2\left[1+c\cos\left(\theta-\alpha\right)\right]},
\label{sol:qinner}
\end{align}
where $c$ and $\alpha$ are constants to be determined. Since the pressure term $\eps/K_{II}$ is negligible in first approximation, we recover, to leading order, the classical Euler elastica of \S~\ref{sec:Euler} with the horizontal direction rotated by an angle $\alpha$. The negative sign of $K_{II}$ will be justified in the next section.  In particular, we may directly infer in the case $c=-1$ that [see Eq.~(\ref{syntractrix})]
\begin{align}
c&=-1:
&z&\sim z_{II}\sim \hat z_0\pm\sqrt{\frac BH} e^{i\alpha}\left(2 e^{i(\theta-\alpha)/2}+\ln\left|4\tan\frac{\theta-\alpha}4\right|\right).
\label{eq:z_II}
\end{align}
In this last expression, $\sqrt{B/H}$ is in fact equal to $\eps H/p$, which makes explicit that the region of space spanned by the large-curvature region is a factor $\eps$ smaller than in low-curvature region. The factor $e^{i\alpha}$ in (\ref{eq:z_II}) indicates that the elastica is rotated by an angle $\alpha$ in the complex plane, and the argument $\theta-\alpha$ in the trigonometric functions come from the fact that the direction of reference for this functional dependence was also rotated by an angle $\alpha$ compared to \S~\ref{sec:Euler}. Alternatively, one can note that $\kappa(\theta)=\kappa_\text{E}(\theta-\alpha)$, where $\kappa_\text{E}(\theta)$ is the curvature derived in \S~\ref{sec:Euler}. Then Eq.~(\ref{elastica:z}) becomes
\begin{equation}
\dd_\theta z=\frac{e^{i\theta}}{\kappa_\text{E}(\theta-\alpha)}=e^{i\alpha}\frac{e^{i(\theta-\alpha)}}{\kappa_\text{E}(\theta-\alpha)},
\end{equation}
and the link between (\ref{syntractrix}) and (\ref{eq:z_II}) directly follows. Finally, the $\pm$ sign in (\ref{eq:z_II}) is deduced as follows: in (\ref{sol:qinner}), $-\sqrt{2[1-\cos(\theta-\alpha)]}=\pm2\sin[(\theta-\alpha)/2]$. Hence, if $\theta<\alpha$, a $+$ sign corresponds to a negative curvature and \textit{vice versa}. 

\subsection{Inflexion point}\label{sec:inflex}
We now examine the vicinity of the inflexion point, where $\kappa$ vanishes, \textit{i.e.},  where $\theta$ reaches an extremum. As a function of $\theta$, the curvature undergoes rapid variations in order to  achieve the connection between the two limiting behaviours, $K_I$ and $\eps^{-1}K_{II}$. Denoting the extremum value of $\theta$ by $\theta_*$, let us write
\begin{align}
K&=K_{III}(\xi), &Q&=Q_{III}(\xi), &\xi&=\frac{\theta-\theta_*}\eps.
\end{align}
Written in this way, the functions $K_{III}$ and $Q_{III}$ vary rapidly in the vicinity of $\theta_*$ in the small-$\eps$ limit. Eq.~(\ref{eq:q2}) then becomes
\beq
\dd_\xi^2 Q_{III}+\frac1{K_{III}}=1+O\left(\eps^2\right).
\eeq
Expressing this equation in terms of $K_{III}$ only yields
\beq
\dd_\xi\left(K_{III}\dd_\xi K_{III}\right)+\frac1{K_{III}}\sim1.
\eeq
Multiplying by $K_{III}\dd_\xi K_{III}$ and integrating, we obtain
\beq
\frac12\left(K_{III}\dd_\xi K_{III}\right)^2+K_{III}=\frac12 K_{III}^2+\mathcal{E}.
\label{eq:bendocap2}
\eeq
where $\mathcal{E}$ is a constant. On the positive-curvature side of the inflexion point, we have the matching condition
\beq
\lim_{\xi\to\pm\infty}K_{III}\sim\lim_{\theta\to\theta_*}K_{I}=1.
\eeq
Above, the limit $\xi\to-\infty$ is implied if $\theta_*$ is a maximum and $\xi\to\infty$ if $\theta_*$ is a minimum. In either case, it follows that $\mathcal{E}=1/2$ in Eq.~(\ref{eq:bendocap2}), which can now be rewritten in terms as
\beq
\left(K_{III}\dd_\xi K_{III}\right)^2=\left(K_{III}-1\right)^2,
\eeq
that is:
\beq
K_{III}\dd_\xi K_{III}=\pm\left( K_{III}-1\right).
\label{eq4KIII}
\eeq
This is easily integrated as
\begin{equation}
e^{K_{III}-1}( K_{III}-1)= -e^{\pm \xi-1},
\label{eq:trans}
\end{equation}
where we used the fact $\xi=0$ is the inflexion point, \textit{i.e.} that $K_{III}(0)=0$. In order for the solution $K_{III}$ of the above equation to be real, it is necessary that $\pm\xi<0$. If the inflexion angle $\theta_*$ is a local maximum, then $K_{III}$ is defined for negative values of $\xi$. Conversely, if   $\theta_*$ is a local minimum, then $K_{III}$ is defined for positive values of $\xi$. Therefore, in the right hand side of Eq.~(\ref{eq:trans}), we must have
\begin{align}
-e^{ \xi-1} & \text{ if $\theta_*$ is maximum,}\nonumber\\
-e^{- \xi-1} & \text{ if $\theta_*$ is minimum.}\nonumber
\end{align}
The transcendental equation $w e^w=z$ possesses two real solutions for $w$, namely the Lambert functions $W_0(z)$ and $W_{-1}(z)$.  Hence, two solutions exist for $K_{III}$. The former is 
\begin{equation}
K_{III}=1+W_0(-e^{\pm\xi-1}),  
\end{equation}
with far field  behaviour
\begin{align}
K_{III}&\sim 1, & Q_{III}&\sim\frac12, &\text{as } \pm\xi\to -\infty.
\label{farfield1}
\end{align}
The latter is
\begin{equation}
K_{III}=1+W_{-1}(-e^{\pm\xi-1}),  
\end{equation}
with
\begin{align}
K_{III}&\sim \pm\xi,
&Q_{III}&\sim \frac{\xi^2}2=\frac{\left(\theta-\theta_*\right)^2}{2\eps^2},
 &\text{as } \pm\xi\to-\infty.
 \label{farfield2}
\end{align}
The two solutions coincide and vanish at the inflexion point $\xi=0$. Moreover, it is immediate to check [compare (\ref{eq:K_I}) and (\ref{farfield1})] that the solution involving the function $W_0$ naturally matches with the low-curvature solution $K_I$. Let us now consider the limiting behaviour of $Q$ in region~$II$ as $\theta\to\theta_*$. Expanding (\ref{sol:qinner}) to second order in $\theta-\theta_*$ and recalling the scaling (\ref{scaling:reg_II}):
\begin{equation}
\eps^{-2}Q_{II}\sim\eps^{-2}\left[ 1+c\cos\left(\theta_*-\alpha\right)-c\sin\left(\theta_*-\alpha\right)\left(\theta-\theta_*\right)
-c\cos\left(\theta_*-\alpha\right)\frac{\left(\theta-\theta_*\right)^2}2\right].
\end{equation}
This matches (\ref{farfield2}) if $\alpha=\theta_*$ and $c=-1$. Furthermore, with  $W_{-1}$, the function $K_{III}$ is negative and, hence, so is the curvature, in the high-curvature region. This justifies the negative sign of $K_{II}$ in (\ref{sol:qinner}).

Finally, let us compute local approximations of the elastic curve in the vicinity of the inflexion point:
\begin{multline}
z
\sim z_{III} 
\sim  z_*+
\frac{H}{p}\int_{\theta_*}^\theta  \frac{e^{i\theta'}\rd\theta'}{K_{III}\left[\left(\theta'-\theta_*\right)/\eps\right]}
\sim z_*+ \frac{\eps H}{p} e^{i\theta_*} \int_{0}^\xi \frac{\rd\xi'}{K_{III}\left(\xi'\right)}
\\
= z_*\pm \frac{\eps H}{p} e^{i\theta_*} \ln\left|1-K_{III}\left(\xi\right)\right|
= z_*\pm \frac{\eps H}{p} e^{i\theta_*} \ln\left|W_{k}(-e^{\pm\xi-1})\right|
,
\label{z_III}
\end{multline}
where we used Eq.~(\ref{eq4KIII}) to perform the integration. Above, $z_*$ is the location of the inflection point and $k=0,-1$ depending on the sign of the curvature.

On the positive-curvature side ($k=0$), we find the far-field behaviour
\begin{align}
K_{III}&>0: 
&z_{III}&\sim  z_*+ \frac{\eps H}{p} e^{i\theta_*} \xi, 
&\pm\xi\to-\infty.
\label{farfield:z_III:low}
\end{align}
On the  negative-curvature side ($k=-1$), we have
\begin{align}
K_{III}&<0: 
&z_{III}&\sim  z_*\pm \frac{\eps H}{p} e^{i\theta_*} \ln\left|\xi\right|,
&\pm\xi\to-\infty.
\label{farfield:z_III:high}
\end{align}

\subsection{Composite approximations}\label{sec:compo}
\begin{figure}\centering
\includegraphics[width=.9\textwidth]{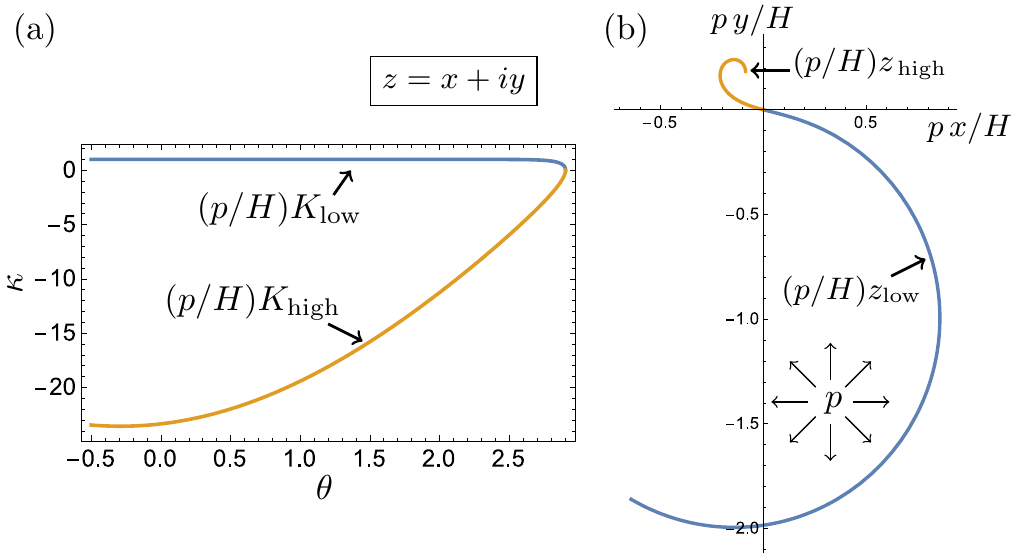}
\caption{Illustration of the composite approximation of the hook shape with $\eps=0.1$ and $\theta_*=2.9$. (a) Curvature  as a function of $\theta$. The blue and orange curves are given by Eq.~(\ref{composite:low}) and Eq.~(\ref{composite:high}), respectively. (b) Corresponding elastica shape in physical space, drawn using Eqs.~(\ref{zlow}) and (\ref{zhigh}). The arrows indicates on which side of the elastica a  net pressure $p$ is applied.}\label{fig:hook}
\end{figure}

Combining the previous approximations, composite, uniformly valid, asymptotic approximations can be constructed (See, e.g., \cite{BObook}). In the entire positive-curvature region, we write the curvature as
\beq
K\sim K_I+ K_{III} - \text{common part},
\eeq
Where the ``common part'' is the function that $K_I$ and $K_{III}$ have in common in the region where both approximations are simultaneously valid. Presently, the common part is simply $1$. Hence, a uniformly valid approximation in the positive-curvature region is
\beq
K_\text{low}\sim1+W_{0}(-e^{\pm\xi-1}).
\label{composite:low}
\eeq
Let us further construct a uniform approximation for $z$. Comparing the asymptotic behaviour of (\ref{eq:z_I}) in the limit $\theta\to\theta_*$ with (\ref{farfield:z_III:low}), we have the matching condition
\beq
z_0-\frac{iH}{p} e^{i\theta_*}\left[1+i\left(\theta-\theta_*\right)\right]
=z_* + \frac{H}{p} e^{i\theta_*} \left( \theta-\theta_*\right).
\label{match:z_I+z_III}
\eeq
This condition can be used to obtain the complex integration constant $z_0$, given $z_*$. The two members of this equality are also the common part of (\ref{eq:z_I}) and  (\ref{z_III}) in their overlapping region of validity. Hence, a uniformly valid, composite expression of the elastica shape in the region of low, positive curvature is obtained by adding the formulas for $z$ in (\ref{eq:z_I}) and  (\ref{z_III}) and subtracting the left hand side of (\ref{match:z_I+z_III}):
\begin{equation}
z_\text{low}\sim\frac{H}{p}\left\{ -i\left(e^{i\theta}-e^{i\theta_*}\right)+e^{i\theta_*}\left[-\theta+\theta_*\pm\eps\ln\left|W_0\left(-e^{\pm\xi-1}\right)\right|\right]\right\}.
\label{zlow}
\end{equation}
In this expression,  we have omitted the constant $z_*$, which amounts to an arbitrary translation in the complex plane.

Turning now to the negative-curvature side of the inflexion point, the common part of $K_{III}$ and $\eps^{-1}K_{II}$ in the region of overlapping validity is $\pm\xi$, see  (\ref{farfield2}). Hence a composite approximation is
\begin{equation}
K_\text{high}\sim1+W_{-1}(-e^{\pm\xi-1})
\pm\eps^{-1}\left[2\sin\left(\frac{\theta-\theta_*}{2}\right)-\theta+\theta_*\right].
\label{composite:high}
\end{equation}
Next, comparing the far field  limit $\theta\to\alpha$ of (\ref{eq:z_II}) with (\ref{farfield:z_III:high}), we obtain the matching condition
\beq
\hat z_0\pm  \frac{\eps H}{p} e^{i\alpha}\left( 2+\ln\left|\theta-\alpha\right|\right)
=z_*\pm \frac{\eps H}{p} e^{i\theta_*} \ln\left|\xi \right|.
\label{match:z_I+z_II}
\eeq
Noting that $\ln\left|\xi \right|=\ln\eps^{-1}+\ln\left|\theta-\theta_* \right|$, we obtain the matching conditions
\begin{align}
\alpha&=\theta_* ,
& \hat z_0\pm  \frac{2\eps H}{p} e^{i\alpha}&
=z_*\pm \frac{\eps H}{p} e^{i\theta_*} \ln\eps^{-1}.
\end{align}
The expression obtained for $\alpha$ is consistent with what we found earlier. The common part of $z_{II}$ and $z_{III}$ in their overlapping region of validity being given by both sides of Eq.~(\ref{match:z_I+z_II}). Adding $z_{II}$ and $z_{III}$ and subtracting the left side of  Eq.~(\ref{match:z_I+z_II}), we obtain the following uniform approximation in the high- negative-curvature region:
\beq
z_\text{\,high}\sim  \pm \frac{\eps H}{p} e^{i\theta_*}\left[4i e^{i(\theta-\theta_*)/4}\sin\frac{\theta-\theta_*}{4}+ \ln\left|\frac{4}{\theta-\theta_* }\tan\frac{\theta-\theta_* }{4}\right|
+\ln\left|W_{-1}\left(-e^{\pm\xi-1}\right)\right|\right],
\label{zhigh}
\eeq
again, up to an arbitrary complex translation $z_*$. 

Equations (\ref{composite:low}), (\ref{zlow}), (\ref{composite:high}), and (\ref{zhigh}) are the main analytical result of this paper and  are illustrated for a given inflexion angle $\theta_*$ and reduced bending stiffness $\eps$ in Fig.~\ref{fig:hook}. 
They form the leading-order description of the elastica in the small-$\eps$ limit and we will indeed see that they can be used to explicitly construct solutions in a wide variety of situations. In practice, Eqs.~(\ref{zlow}) and (\ref{zhigh}) yield curves that are remarkably close to the actual solutions of the elastica solution even for moderately small values of $\eps$. We note however that, away from the inflexion point, the actual angle made by the tangent to curves drawn with these formula differs from $\theta$ by an $O(\eps)$ quantity. In other words, the parameter $\theta$ appearing in  Eqs.~(\ref{zlow}) and (\ref{zhigh}) is not exactly the angle made by the tangent of these curves with the horizontal.

In response to this issue, an alternative to   Eqs.~(\ref{zlow}) and (\ref{zhigh}) is to numerically integrate
\begin{align}
z&\sim\frac{H}{p}\int_{\theta_*}^\theta\frac{e^{i\theta'}}{K(\theta')}\rd \theta',
&K&=K_\text{low}, K_\text{\,high}.
\label{z:alt}
\end{align}
We will use this expression to compute the condition of self-contact in the next section.

\section{Application 3: pinching a ring into a peanut or a trefoil shape}\label{sec:peanut}


\subsection{Experiments and construction of the shape}
In order to assess the validity of this asymptotic theory, we carry experiments and numerical simulations where an elastic ring of radius $R$ and square cross section of side $h$ is loaded with an internal pressure $p$ and perturbed with two diametrically opposed indenters. 

We manufacture our elastic rings by casting a silicon elastomer (Elite Double 8 from Zhermack, Young Modulus $E=250\pm15$ kPa) into 3D-printed moulds. These elastic rings are then placed on the surface of a bath of distilled water. For each experiment, we initially add a small quantity of dishsoap at the surface of the bath inside the ring. This results in a sudden drop in the surface tension inside the ring. The difference of surface tension on either side of the elastica produces a net outward force $p$ per unit length on the order of 0.04Nm$^{-1}$~\cite{Pineirua2013}. The elastica is additionally constrained by two opposite indenters fixed at a given distance $d$ from each other, imposing point forces on the structure. Given the side length $h$, $B=Eh^4/12$ and we have the dimensionless bendability 
\beq
\epsilon_0=\sqrt{\frac{B}{pR^3}}.
\eeq
We will see shortly [Eq.~(\ref{eps_vs_theta1})] that $\eps_0$ is closely related to $\eps$. Compared to the latter, it has the advantage of being fixed for a given experimental set-up, independently of deformation. By varying $h$ in the range $[0.5,2]$\,mm and $R$ in the range $[24,53]$\,mm, we are able to experimentally achieve $0.01<\eps_0<1.1$. Note that in their study of dynamical buckling, Kodio {\it et al}.~\cite{Kodio2020} used the equivalent dimensionless parameter $P=1/\eps_0^2$. Their linear stability analysis in the large-$P$ limit showed that the most unstable mode number scales as $(P/2)^{1/2}$, suggesting the excitation of boundary layers by the instability.

\begin{figure}
\centering
\includegraphics[width=.9\textwidth]{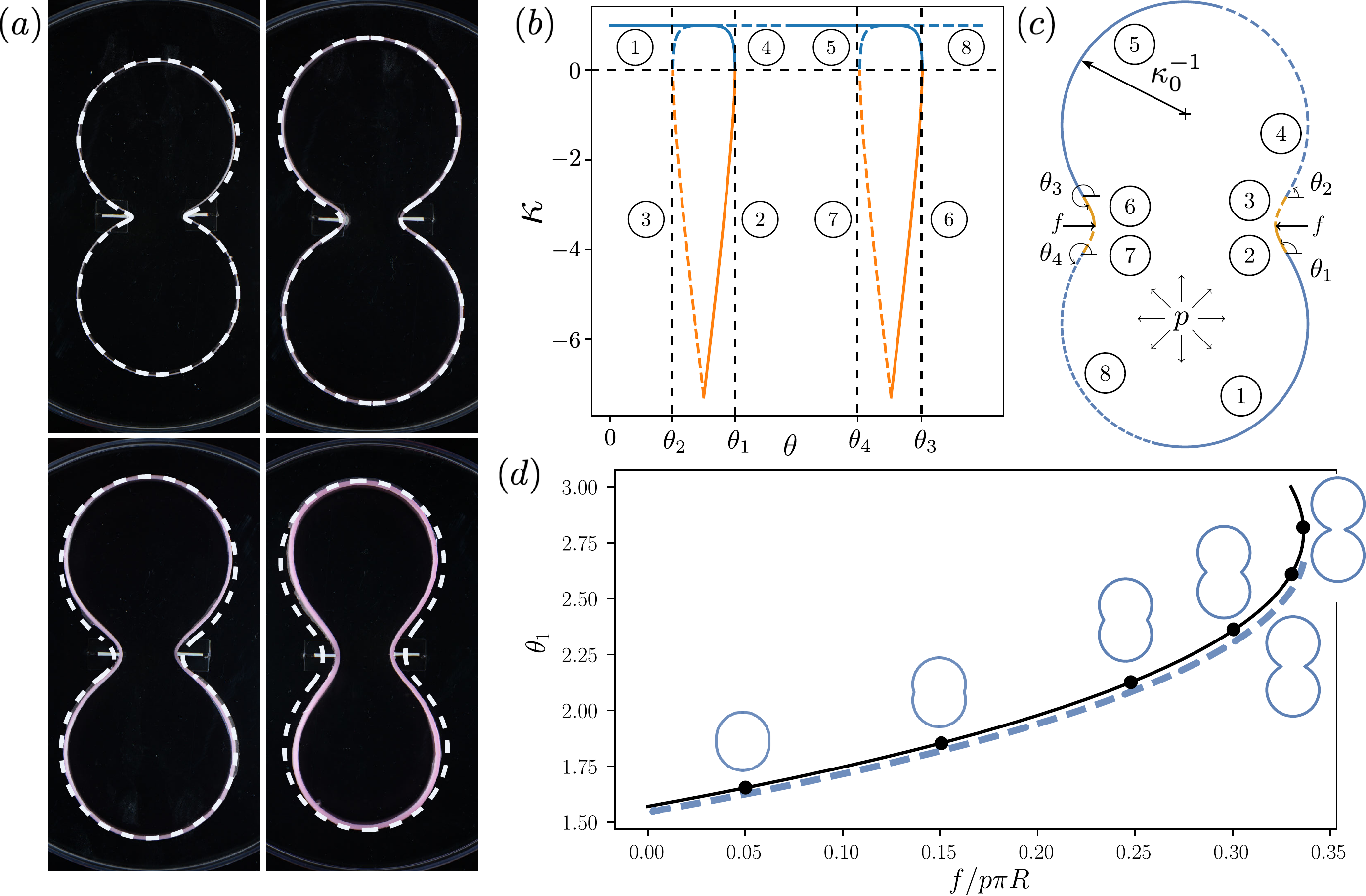}
\caption{(a) Comparison between experimental and analytical shapes of the peanut for increasing values of $\eps_0$ ($h=0.4,0.8,1,1.4$ mm , $R=47.2,53.4,51.5,47.1$ mm, corresponding to $\eps_0=0.011,0.037,0.061,0.138$ respectively; $d=19.2$ mm). The value of the inflexion angle $\theta_1$ required to compute the shape is taken from the numerical simulation with the same parameters. 
(b) $\kappa$ vs $\theta$ in an elastic circular ring subjected to outward pressure and two opposite inward point forces. $\eps=0.1$. (c) the resulting peanut shape, constructed  piecewise using Eqs.~(\ref{composite:low}), (\ref{zlow}), (\ref{composite:high}), and (\ref{zhigh}). The circled numbers indicate correspondance between the two graphs. (d) Inflexion angle $\theta_1$ as a function of the rescaled point force $f/p\pi R$. The solid line corresponds to the theory, see Eq.~(\ref{peanutforce}), the dashed line to the numerical simulations with  $\eps_0=5.2\times10^{-3}$ . Beyond the rightmost point, the elastica snaps to contact. Animations showing deformation and snapping are given in Supplementary Information.
}\label{fig:peanuta}
\end{figure}

For small values of  $\epsilon_0$, this setup leads to peanuts shapes with two positive, nearly constant,  low-curvature regions separated by two boundary layers of large negative curvature near the indenters point where the two forces are applied. As $\epsilon_0$ is increased, these boundary layers expand [Fig.~\ref{fig:peanuta}(a)].

Such shapes can be asymptotically described by combining solutions of the forms (\ref{zlow}) and (\ref{zhigh}) derived in the previous section.  A point force at a given arc length $s_0$ is modelled by the addition of a delta function in the right hand side of  Eq.~(\ref{eq:nperp1}): $p\to p-f\delta\left(s-s_0\right)$. This produces the discontinuities
\beq
\left[B\dd_s\kappa\right]^+_{-}
=-\left[\n\right]^+_{-}=f,
\eeq
In the $\theta$-formulation, the jump is
\begin{align}
\left[\dd_\theta q\right]^+_{-}&=f/B, 
&\iff&
&\left[\dd_\theta Q\right]^+_{-}&=\frac{f}{H\eps^2}. \label{discont}
\end{align}
Bearing this discontinuity in mind, it is simple to stitch the solutions (\ref{composite:low}) and (\ref{composite:high}) into a piecewise function that approximates the curvature along the deformed ring. This is illustrated in Fig.~\ref{fig:peanuta}. In the case of a peanut shape, there are 4 inflection points with angles $\theta_1$ to $\theta_4$ as defined in the Fig.~\ref{fig:peanuta}(c). By symmetry, only $\theta_1$  needs to be specified, since
\begin{align}
\theta_2&=\pi-\theta_1, &\theta_3&=\pi+\theta_1, &\theta_4&=2\pi-\theta_1.
\end{align}
The solution can be constructed by substituting $\theta_*=\theta_1$, $\theta_2$,  $\theta_3$, and $\theta_4$ and adding appropriate complex translations in the formulas of \S~\ref{sec:hook} (See Mathematica files as Supplementary Material for  examples of implementation). Given $\eps$, it appears from these formulas that the shape is entirely determined, up to a scale factor $H/p$, by $\theta_1$. Fig.~\ref{fig:peanuta}(b) and (c) displays the result obtained by evaluating Eqs.~(\ref{composite:low}) and  (\ref{composite:high}) and Eqs.~(\ref{zlow}), and (\ref{zhigh}), respectively, for $\eps=0.1$ and $\theta_1=2.9$. It is visually clear that the existence of an inflexion point requires $\theta_1>\pi/2$, while selfcontact necessarily arises as $\theta_1\to\pi$ in the $\eps\to0$ limit.
 
The slope discontinuities in Fig.~\ref{fig:peanuta}(b) at $\theta=\pi/2$ and $3\pi/2$ are due to the point forces. Let us now evaluate them. On either side of $\theta=\pi/2$, in the vicinity of the applied force, we have $Q_{-}\sim\left[1-\cos\left(\theta-\theta_1\right)\right]/\eps^2$ and $Q_{+}\sim \left[1-\cos\left(\theta-\theta_2\right)\right]/\eps^2$, respectively. Evaluating expression (\ref{discont}) at $\theta=\pi/2$, we thus find
\beq
f=-2H\cos\theta_1.
\label{discont2}
\eeq
Meanwhile, the large-absolute curvature regions of the ring only make up an $O(\eps)$ fraction of its length. Hence, the length of the deformed ring is, to leading order, given by $4\theta_1/\kappa_0$, where $\kappa_0$ is the nearly constant curvature away from the pinch. If the ring is inextensible, this length must be equal to $2\pi R$:
\beq
\frac{4\theta_1}{\kappa_0}\sim 2\pi R+O\left(\eps\right).
\eeq
Now $\kappa_0$ corresponds to $K\sim1$, so $\kappa_0\sim p/H$. Hence, $H\sim  \pi p R/(2\theta_1)$. Substituting this expression into Eq.~(\ref{discont2}), we eventually obtain
\begin{align}
\text{peanut:}&&f&\sim -p\pi R  \frac{\cos\theta_1}{\theta_1} +O\left(\eps\right),
&\pi/2\leq\theta_1<\pi.
\label{peanutforce}
\end{align}
This relates the applied point force to the resulting inflexion angle $\theta_1$, \textit{i.e.}, to the deformation of the ring. Using the expression of $H$ that we just derived, we further have
\begin{align}
\eps^2 &= \frac{B}{ p R^3}\left(\frac{2\theta_1}\pi\right)^3\equiv \eps_0^2 \left(\frac{2\theta_1}\pi\right)^3,
&\eps_0&=\sqrt{\frac{B}{ p R^3}}.
\label{eps_vs_theta1}
\end{align}
where the bendability parameter $\eps_0$, introduced earlier in this section, is the value of $\eps$ obtained in the limit $\theta_1\to\pi/2$, \textit{i.e.}, $f\to0$. 
It is important to realise that the small parameter $\eps$ on which the asymptotic development is based depends on $H$ and, hence, varies in the course of deformation, see Eq.~(\ref{eps_vs_theta1}). While $\eps$ is more natural and convenient from an analytical point of view, the alternative small parameter $\eps_0$ is more sensible from an experimental point of view. Indeed, its value is fixed once and for all given $B$, $p$ and $R$ and independently of the point force $f$.

Taking the value of the inflection angle $\theta_1$ given by numerical simulations (detailed below), we may thus reconstruct the shape of the peanuts without any fitting parameter [dashed lines in Fig.~\ref{fig:peanuta}(a)]. As expected, we obtain a very good agreement between experiments and theory in the small $\epsilon_0$ limit up to $\epsilon_0\approx0.1$, past which the accuracy of the asymptotic theory gradually decreases.

Finally, evaluating the length $\ell$ of the peanut shape on the basis of the outer solution, we easily derive, in the limit $\eps\to0$,
\beq
\ell(\theta_1)=\frac{2H}p\left(1-\cos\theta_1\right)=\pi R\frac{1-\cos\theta_1}{\theta_1}.
\eeq
The surprising feature of this formula is that $\ell(\theta_1)$ passes by a maximum value of $\ell(2.33)\approx2.28R$ in the course of deformation, with  $\ell(\pi)=\ell(\pi/2)=2R$. This follows entirely from the inextensibility constraint.

\subsection{Snap-through buckling}

Equation~(\ref{peanutforce}) implies, unexpectedly, that there is a maximum value of $f$ beyond which there is no solution, see Fig.~\ref{fig:peanuta}(d). This limit point (LP) corresponds to  a value of $\theta_1$ given by $\theta_{1,LP}\approx2.80$, where  $f=f_{LP} \approx0.34p\pi R$. To explain this observation, note that, at the inflexion point, $H$ is exactly given by $\npa$. Therefore, with the aid of Fig.~\ref{fig:peanuta}, it is immediate to recognize in (\ref{discont2}) the horizontal force balance 
\beq
f\sim-2\npa\cos\theta_1
\eeq
over the entire negative-curvature region. Indeed, recalling that its extent is only of $O\left(\eps\right)$, the pressure contribution is negligible to leading order in that equilibrium condition. The absence of a static solution for $f>f_{LP}$ can thus be simply understood by the fact that the tension $\npa$ inside the ring is not sufficient to balance the point force $f$. Henceforth, we conclude that the ring must snap into self-contact. Controlling the force rather than the displacement in the experiments is challenging, and we therefore turn to numerical simulations to check this prediction.

We simulate the ring by discrete  Cosserat rod equations, which we integrate using the method of Ref.~\cite{Gazzola2018}. In this scheme, rotation matrices are output for every discrete point of the elastica and the angle $\theta(s)$ can directly be retrieved from it. We study separately displacement-controlled and force-controlled deformations.

In the first scenario, we impose the distance $d$ between two diametrically opposed points of the ring. We use values of $B$, $R$ and $p$ directly inferred from the experimental parameters, which fixes the value of $\eps_0$. Starting from the unstrained state $d=2R$, we quasistatically decrease $d$ and monitor the inflexion point $\theta_1$ as a function of $d$. This value of $\theta_1$ is then substituted in the analytical formulas to compare analytical and experimental curves in Fig.~\ref{fig:peanuta}(a).

In the second scenario, the value of the force $f$ is imposed and $\theta_1$ and $d$ are monitored as a function of $f$. For very small values of $\eps_0$, self-contact arises, as predicted, by a snap-through at the limit point value $f_{LP}$ and very good agreement is found with the analytical formula, Eq.~(\ref{peanutforce}), as shown in Fig.~\ref{fig:peanuta}(d). As $\eps_0$ is increased, the influence of the boundary layer becomes more important. For $\eps_0>0.13$, the deformation in the boundary layers is sufficiently large that  self-contact happens for $f<f_{LP}$. In that case, no snap-through buckling is observed, and the two scenarii (displacement-controlled and force-controlled) become equivalent.

\begin{figure}
\centering
\includegraphics[width=\textwidth]{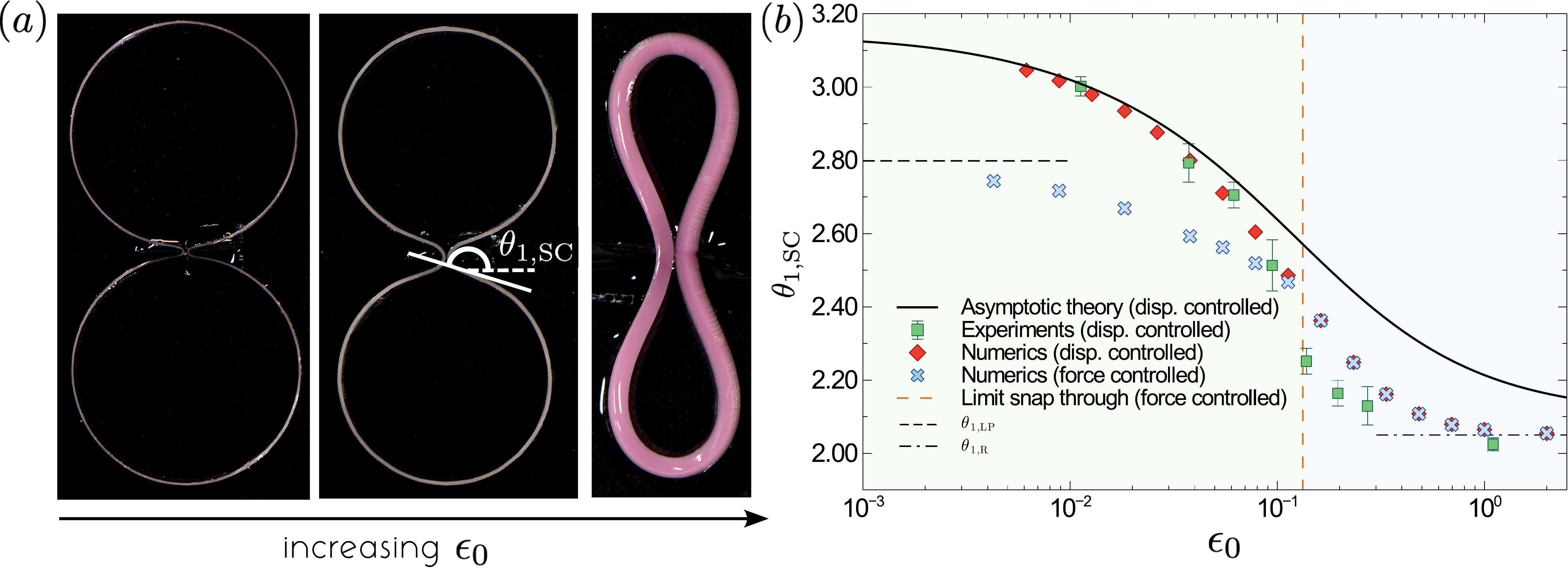}
\caption{(a) Pictures of experimental peanuts shape when selfcontact is imposed for increasing values of $\eps_0$ ($\eps_0=0.01$, 0.04 and 1.1 respectively). (b) Critical angle $\theta_{1,\text{SC}}$ at the transition to selfcontact as a function of $\eps_0$ for the theory, numerical simulations and experiments. }\label{fig:force}.
\end{figure}

\subsection{Transition to selfcontact}
In order to quantitatively assess the regime of validity of the asymptotic theory, we enquire under which condition self-contact arises in both scenarios (force- or displacement-controlled).

As mentioned before, when the control parameter is the applied force $f$, the leading-order theory as $\eps_0\to0$ predicts a critical inflexion angle $\theta_{1,LP}\approx2.80$ at which the elastica snaps to contact. As $\epsilon_0$ is increased from zero, however, selfcontact can occur before this critical angle is reached. In such a case, controlling the displacement or the force are thus equivalent.

The condition of self-contact is simply expressed by saying that $x(\theta=\pi/2)$ in the negative-curvature region is equal to $x(\theta=\pi/2)$ in the positive curvature region. Using Eq.~(\ref{z:alt}), the condition is
 \beq
 \int_0^{\theta_1}\frac{\cos{\theta}}{K_\text{low}(\theta,\eps)}\rd\theta= \int_{\pi/2}^{\theta_1}\frac{\cos{\theta}}{K_\text{\,high}(\theta,\eps)}\rd\theta.
 \label{eq:contactCondition}
 \eeq
This equation yields the value of $\theta_{1,\text{SC}}$ at which self-contact happens as a function of $\eps$.  Solving Eq.~(\ref{eq:contactCondition}) as an equation for $\theta_1$ with different values of $\eps$, we eventually obtain
\begin{align}
\text{self-contact:}
&& \theta_{1,\text{SC}}\approx\pi\times\frac{1+189\eps+474\eps^2}{1+192\eps+722\eps^2}\approx
\pi\times\frac{1+103\eps_0+522\eps_0^2}{1+109\eps_0+778\eps_0^2},
\label{theta_SC_predicted}
\end{align}
where we recall that $\eps_0=\eps/(2\theta_1/\pi)^{3/2}$.
For instance, with $\eps_0=0.1$, we find  $\theta_1\approx2.64$ at self-contact and estimate through Eq.~(\ref{peanutforce}) that this happens for an applied force $f\approx0.332\times p\pi R$.

In Fig.~\ref{fig:force}, we compare the above analytical estimate with the experimental values for displacement-controlled deformation and numerical simulations for both displacement- and force-controlled deformation. The value $\eps_0^*\approx0.13$ marks the separation between numerical outcomes when either $d$ or $f$ is imposed in the numerical code. For $\eps_0<0.13$, snap-through is numerically observed and the inflexion angle corresponding to this instability is seen to tend to the analytically predicted value as $\eps_0$ is further decreased.  An example of  snap-through dynamic observed in our numerical simulations for a fixed force $f$ slightly larger than the critical value is shown in the supplementary movie for $\epsilon_0 = 9\times 10^{-3}$. 

Overall, a good agreement is found between the three approaches for small enough $\eps_0$. For large $\eps_0$, the quantitative predictions of the boundary layer theory become inaccurate. Both numerical simulations and experiments indicate a plateau in Fig.~\ref{fig:force}(b). For such large values of $\eps_0$ and, hence, $\eps$, the pressure term in Eq.~(\ref{eq:q1}) becomes irrelevant and we recover the classical problem of \S~\ref{sec:Euler}. In that frame, we note that selfcontact arises at  $c=2.16279$ (Fig.~\ref{fig:Euler}). Using this  value in Eq.~(\ref{euler:q}), we find an inflexion angle $\arccos(-1/2.16279)\approx2.05$ (dashed black line to the right in Fig. \ref{fig:force}b). Such a situation also corresponds to the racket shape obtained experimentally by Py {\it et al.} in Ref.~\cite{Py07}.

\begin{figure}
\centering
\includegraphics[width=0.8\linewidth]{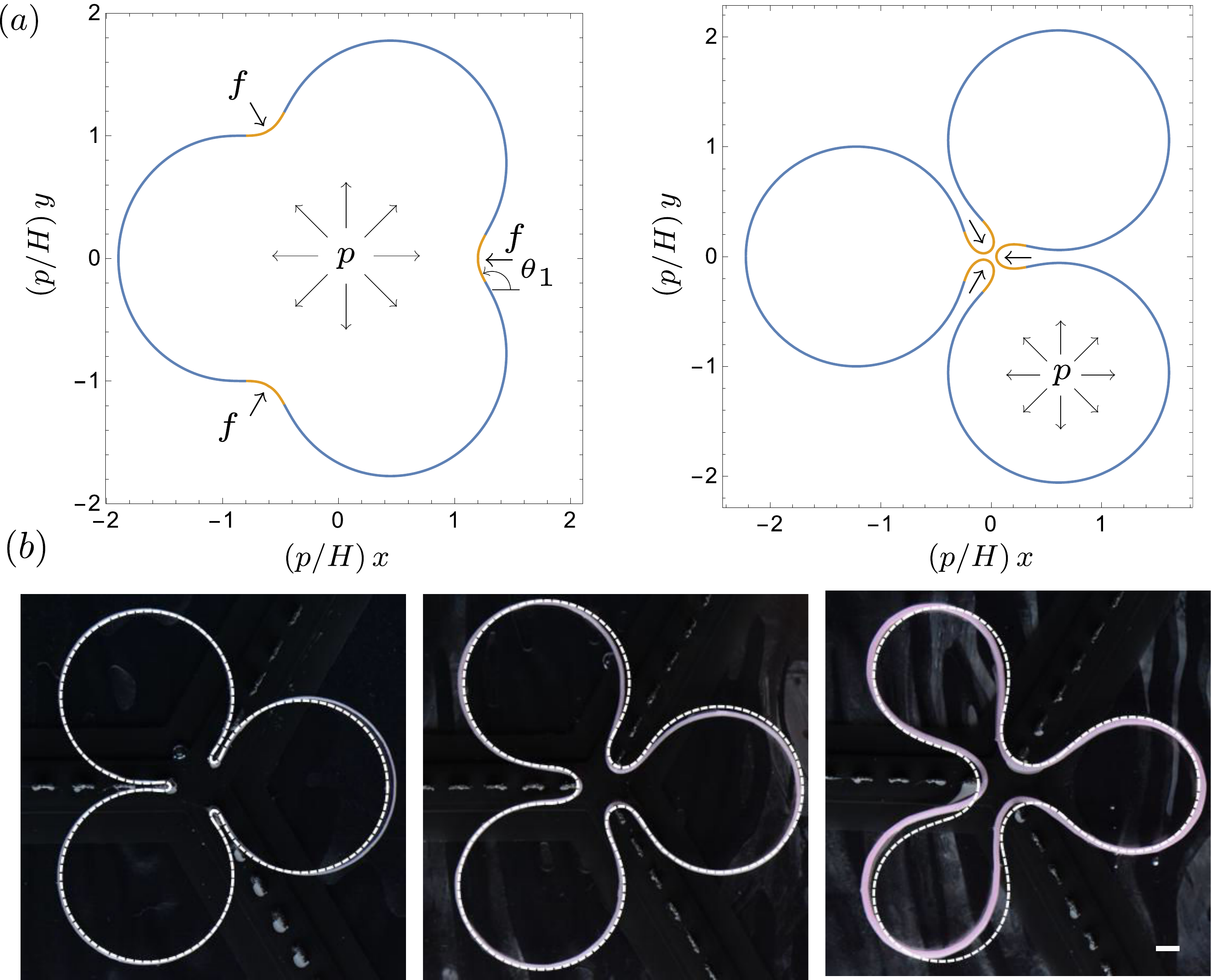}
\caption{(a) An elastic circular ring deformed into a trefoil by three opposite point forces. Left: $\eps=0.1$ and $\theta_1=2\pi/3$. Right: $\eps=0.1$ and $\theta_1=3.34$. The latter configuration is mechanically unstable if the control parameter is $f$ (see text). (b) Comparison between experimental and theoretical trefoil shapes for increasing values of $\epsilon_0$ ($\epsilon_0=0.01$, 0.04, 0.09).}.\label{fig:trefoil1}
\end{figure}
\subsection{Trefoil and $N$-foil shape}
Thanks to the simplicity brought about by the preceding asymptotic argument, we can easily generalize our discussion to the case of $N>2$ point forces applied symmetrically to an elastic ring. The relation (\ref{discont2}) continues to hold on each side of the applied point force but this time, the inextensibility condition is
\beq
\frac{2N\theta_1}{\kappa_0}\sim 2\pi R+O\left(\eps\right).
\eeq
Hence, for a symmetric $N$-foil shape, we have
\begin{align}
\text{$N$-foil :}&&f_N&\sim -p\pi R  \frac{2\cos\theta_1}{N\theta_1} +O\left(\eps\right),
&\eps&=\eps_0\left(\frac{N\theta_1}{\pi}\right)^{3/2},
\label{Nfoilforce}
\end{align}
where, as before, $\theta_1=\pi/2$ correspond to a vanishing compressing force applied horizontally and $\eps_0=\sqrt{B/pR^3}$. The maximum force for which a solution exists decreases as $1/N$ but the maximum inflexion angle remains unchanged. The symmetric trefoil is illustrated for two values of $\eps$ and $\theta_1$ in Figs.~\ref{fig:trefoil1}. (Mathematica file to generate the figure given in Supplementary Information.) One of the situation depicted correspond to $\theta_1>\theta_{1,LP}$ and is therefore mechanically unstable in the case of a prescribed value of $f$. Again, if the displacement at the pinching point is prescribed, rather than $f$, no instability occurs. Furthermore, we note that, contrary to the peanut, a nonzero value of $\eps$ allows one to attain values of $\theta_1$ larger than $\pi$. The existence of a limit point again implies a snapping instability in the force-driven deformation and this is again confirmed by numerical simulations. It is remarkable that the LP remains associated to the same value of $\theta_1$, independently of $N$ and $p$.

\paragraph*{\bf Remark}
So far, we have confined our attention to symmetrically applied forces, as this automatically ensures static equilibrium. More general sets of forces can obviously be applied, provided that they are again in static equilibrium, \textit{i.e.}, their vector sum and total moment both vanish. One technical difference applies in that case: the tension $\npa$ also becomes discontinuous at the points where loads are applied. Indeed the applied forces now generally contain a non-vanishing component along the tangent of the elastica. Calling this component of force $f_t$, we have the discontinuous jump
\beq
\left[\npa\right]_{-}^{+}
=-f_t.
\eeq
This results in a discontinuity
\beq
\left[H\right]_{-}^{+}=-f_t
\eeq
in the constant of integration to be used on either side of the point force. Bearing this modification in mind, the same piecewise construction as before, based on Eqs.~(\ref{composite:low}) and (\ref{composite:high}) can in principle be used.

\section{Application 4: the stretched ring}

\begin{figure}
\centering
\includegraphics[width=\linewidth]{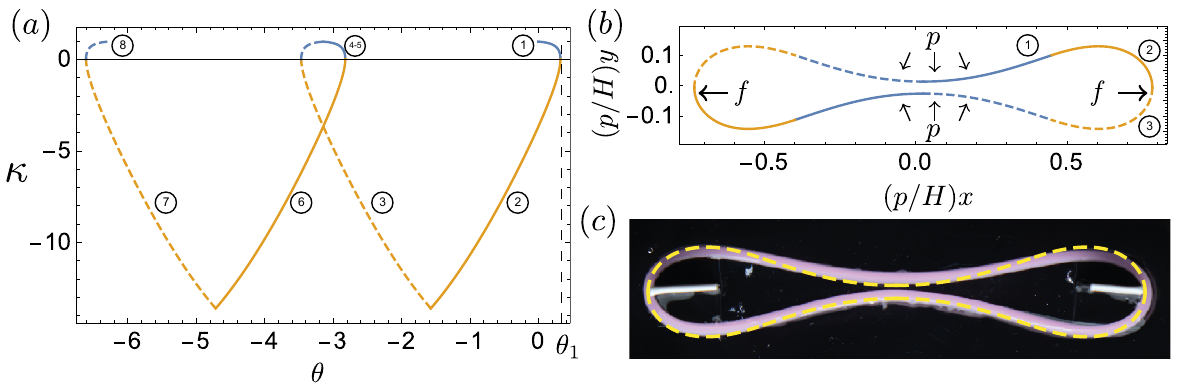}
\caption{Elastic circular ring subjected to inward pressure and two opposite outward point forces with $\eps=0.15$ and $\theta_1=0.32$. (a) Top: $\kappa$ vs $\theta$. (b) Bottom: piecewise analytically computed profile. (c) Comparison with the experiment: dishsoap is added to the bath on the outside of the elastic ring.}\label{fig:antipeanut}
\end{figure}

Let the pressure now be applied externally on the ring, while two opposite point forces are applied outwards. This is the mechanical inverse of the situation leading to the peanut shape, so we might call it the `anti-peanut'. The formalism derived in previous sections directly applies here too and is depicted in Fig.~\ref{fig:antipeanut}. Interestingly, the piecewise construction of the function $\kappa(\theta)$ yields a qualitatively different picture than with the peanut:  branches of solutions intersect in the large-curvature side in Fig.~\ref{fig:antipeanut} instead of the low-curvature side in Fig.~\ref{fig:peanuta}(b). This is due to the fact that the range of values of $\theta$ where $Q\gg1$ makes up the majority of the domain of $\kappa(\theta)$ in the anti-peanut configuration.

As before, we may capitalize on previous calculation to  directly infer that $f\sim 2\npa\cos\theta_1$ and  that $\kappa t\sim p$ in the low-curvature region, with the same inextensibility constraint as before. We then immediately obtain

\begin{align}
f&\sim p\pi R  \frac{\cos\theta_1}{\theta_1} +O\left(\eps\right),
&0<\theta_1.
\label{antipeanutforce}
\end{align}
This formula for $f$ is consistent with intuition: flatter shapes correspond to smaller $\theta_1$ and hence to larger stretching forces. Above, there is  a maximum allowed value of $\theta_1$ to prevent self-crossing of the two segments of low curvature. The maximum allowed value of $\theta_1$ increases with $\eps$ as follows. In the low-curvature region ($K\sim1$), the curvature is nearly constant, so $\theta(s)\sim\kappa_0 s$. At the inflexion point $\theta_1\sim\kappa_0 s_1$. Meanwhile, the vertical difference between the center of the stretched elastica and the inflexion point is approximately $\kappa_0s_1^2/2=\theta_1^2/(2\kappa_0)$. To prevent self-crossing this vertical elevation must be met by an equivalent opposite vertical displacement in the region of large curvature (see~Fig.~\ref{fig:antipeanut}), the latter being on the order of $\eps/\kappa_0$. Hence, the maximum allowed value of $\theta_1$ is $O\left(\eps^{1/2}\right)$.

At this stage, it is clear that the above configuration can easily be generalized to a larger number of stretching forces, just as in the previous section.

\section{Generalization with body forces}\label{sec:bodyforces}
The success of Eq.~(\ref{eq:q1}) to address complex elastica shape was due the reduction of the original problem by two differential orders: one thanks to the existence of the first integral $H$, the other by switching the curve parameterization to $\theta$. In the presence of body forces, $H$ is not constant anymore but an asymptotic treatment along the previous lines can still be envisaged.  Eqs.~(\ref{eq:nperp1}) to (\ref{eq:kappa1})  become
\begin{align}
\dd_s\n&=p-\kappa\npa-\g_n, 
&\dd_s\npa&=\kappa\n-\g_t, 
&B\dd_s\kappa&=-\n. \label{eq:kappa2}
\end{align}
where $\g_t$ and $\g_n$ are external axial and shear force per unit length, respectively. Recently, for instance, an analytically solvable model for elastic wrinkling on a nonlinear foundation was introduced, in which $\g_t=0$ while $\g_n$ scales as some power of the local displacement of the elastica~\cite{Foster2022a,Foster2022,Foster2022c} (See also Refs.~\cite{Brau2010,Diamant2011,Gordillo2019,Michaels2019,Michaels2021} for more experimental and theoretical results on the subject.)

We may continue to define $H$ as $B\kappa^2/2+\npa$. It now  satisfies
\beq
\dd_s H=-\g_t.
\eeq
In the $\theta$-formulation, we have
\begin{align}
B\left(\dd^2_{\theta}q+q\right)+\frac{p-\g_n}{\kappa}&= H, 
&\dd_\theta H&=\frac{-\g_t}{\kappa}. \label{eq:q4}
\end{align}
In regions of large curvature, we continue to benefit from the simple leading order problem
\begin{align}
B\left(\dd^2_{\theta}q+q\right)&\sim H, 
&H&\sim\text{const.},
\end{align}
and we recover the approximation of previous sections. This time, however, the value of $H$ generally varies between  boundary layers. On the other hand, if  $\kappa$ is `small',  we have $H\sim t$ and  we may  neglect terms proportional to $B$ in Eqs.~(\ref{eq:q4}). We thus have
\begin{align}
\kappa&\sim\frac{p-\g_n}{H}, 
&\frac1H\dd_\theta H&\sim\frac{-\g_t}{p-\g_n} . 
\end{align}
Suppose for instance that gravity acts in the vertical direction, so that $\g_n=-\rho g\cos\theta$, $\g_t=-\rho g\sin\theta$, where $\rho$ is the mass per unit length. The equation for $H$ above is easily solved in that case. We have
\beq
H\sim\frac{\rho g h}{a+\cos\theta},
\eeq
for some constant $h$, where we define
\beq
a=\frac{p}{\rho g}.
\eeq
Meanwhile $\kappa\sim\left(a+\cos\theta\right)^2/h$. Then, integration of Eqs.~(\ref{elastica:xy}) yields, up to an arbitrary translation,
\begin{align}
x&\sim \frac{2h}{\left(1-a^2\right)^{3/2}}\arctanh\left(\frac{1-a}{\sqrt{1-a^2}}\tan\frac{\theta}{2}\right)
-\frac{ah\sin\theta}{\left(1-a^2\right)\left(a+\cos\theta\right)},
&y&\sim \frac{h}{a+\cos\theta}.
\end{align}
Finally near an inflexion point, the same local analysis as in Sec.~\ref{sec:inflex} applies to leading order.

\section{Conclusions and perspectives}

The present  boundary layer theory leads to a composite solution that can be used as a  building-block to describe a wide variety of complex shapes. With $\theta$ as the curve parameter, the solution takes an elementary form. The only concession to the use of special functions is in the vicinity of inflexion points, where the Lambert functions $W_0$ and $W_1$ arise.  Thanks to this simplified picture, simple local expressions of force balance can be derived, allowing intuitive mechanical analysis. This gives confidence, in particular, in the surprising prediction of snapping instability of pinched peanuts and $N$-foil shapes past a certain applied force. It would be interesting to see if such a scenario could find application in engineering or in microbiology. Furthermore, a good agreement was obtained both with direct numerical simulations and experiments.

Not all possible analyses afforded by the simple $\theta$-formulation have been exposed here. First, it is in principle not difficult to compute the first corrections to the leading order expressions that we have derived to obtain more accurate formula and enlarge the range of values of the small parameter $\eps$ where they are reliable. Secondly, not all possible shapes have been envisaged within the present framework. We have already mentioned how contact points with friction, \textit{i.e.}, with a non vanishing tangent component, can be handled. Another situation that we haven't explored is the possibility to have segments of large, opposite curvature in succession, rather than the low-high curvature detailed in Sec.~\ref{sec:hook}. Let us  briefly explain how this could in principle be handled. One would need to match two regions of large curvatures with opposite signs, each described asymptotically by expressions of the form~(\ref{sol:qinner}). The intermediate region is now still described as in Sec.~\ref{sec:inflex}, only this time the constant of integration  $\mathcal{E}$ appearing in~(\ref{eq:bendocap2}) differs from $1/2$. This equation can still be integrated implicitly, yielding $\xi$ as a function of $K_{III}$. Setting $\mathcal{E}=1/2+b^2$, one can derive
\beq
\sqrt{b^2+\left(K_{III}-1\right)^2}+\arcsinh\left(\frac{K_{III}-1}{b}\right)=\pm \xi+c
\eeq
where $c$ is a constant, but the  inversion of the resulting formula is not straightforward. In the limit $\xi=\pm\infty$, one obtains the desired behaviour $K_{III}\sim\pm\xi$. One limit in which the above expression can be inverted asymptotically is $b\to\infty$. We do not pursue the analysis here but note that this limit could be relevant in situations where  two point forces of opposite signs are applied at close locations of the elastica rather than far appart.

\acknowledgments
G.K. is a Research Associate of the Fonds de la Recherche Scientifique - FNRS (Belgium.) This project has received funding from the European Union's Horizon 2020 research and innovation programme under the Marie Sklodowska-Curie grant agreement n$^{\circ}$ 101027862, from the F.R.S.-FNRS under the research grant n$^{\circ}$ J.0017.21 (CDR ``FASTER'') and from the Federation Wallonia-Brussels (Concerted Research Actions ``Capture'').


\vskip2pc

\end{document}